\documentclass[12pt]{iopart}
\usepackage{graphicx,amssymb}

\newcommand{\beq}{\begin{equation}}
\newcommand{\eeq}{\end{equation}}
\newcommand{\beqn}{\begin{eqnarray}}
\newcommand{\eeqn}{\end{eqnarray}}          
\newcommand{\bra}{\left<}

\newcommand{\ket}{\right>}

\newcommand{\bfig}{\begin{figure}}
\newcommand{\efig}{\end{figure}}
\newcommand{\bmp}{\begin{minipage}} 
\newcommand{\emp}{\end{minipage}}
\newcommand{\bc}{\begin{center}}
\newcommand{\ec}{\end{center}}

\begin{document}
\title{
Coherent  control of an effective two-level system  in a non-Markovian  biomolecular  environment 
}
\author{J Eckel$^{1,2}$, J H Reina$^2$ and M Thorwart$^{1,3}$}
\address{$^1$ Institut f\"ur Theoretische Physik, Heinrich-Heine-Universit\"at D\"usseldorf, Universit\"atsstr.1, 40225 D\"usseldorf, Germany}
\address{$^2$ Departamento de F\'isica, Universidad del Valle, A.A. 25360, Cali, Colombia}
\address{$^3$ Freiburg Institute for Advanced Studies (FRIAS), Universit\"at Freiburg, 79104 Freiburg, Germany}
\ead{eckelj@thphy.uni-duesseldorf.de}
\begin{abstract}
We investigate the quantum coherent dynamics of an externally driven effective two-level
system subjected to a slow Ohmic environment characteristic of biomolecular protein-solvent   
reservoirs in  photosynthetic light harvesting complexes. By means of the numerically exact
quasi adiabatic propagator path integral (QUAPI) method we are able to include
non-Markovian features of the environment and show the dependence of the
quantum coherence on the characteristic bath cut-off frequency
$\omega_c$ as well as on the driving frequency $\omega_l$ and the field amplitude $A$. 
Our calculations extend from the weak coupling regime to the incoherent
strong coupling regime. In the latter case, we find evidence for a resonant behaviour,
beyond the expected behaviour, when the reorganization energy $E_r$ coincides 
with the driving frequency. 
Moreover, we investigate how the coherent destruction of tunneling within
the two-level system is influenced by the non-Markovian environment.
\end{abstract} 
\pacs{03.65.Yz,
42.50.Dv,
87.15.H-,
03.67.-a 
}
\submitto{\NJP}

\section{Introduction}  

During the past decade, a tremendous advancement in realizing and controlling
the coherent quantum dynamics in solid-state nanosystems by an external time-dependent force
has been achieved \cite{fazio05}.  The theoretical description of driven quantum systems has discerned novel effects related to the control of quantum tunneling (see, e.g., \cite{grifoni98} for a comprehensive review). 
Classical examples of the achieved degree of quantum control are the manipulation of  trapped atoms in quantum optics \cite{nobel97} and the control of chemical reactions by external 
laser fields \cite{grifoni98}. In quantum optics,  it has been experimentally demonstrated that the
time evolution of a two-level atom can be significantly modified by means of a frequency modulated excitation
of the atom by use of a microwave field \cite{noel98}. Moreover, it has been shown that tunneling of an initially localized state in a double well potential can be almost completely suppressed  by a properly tailored external driving force \cite{grossmann91}. In addition, the experimental demonstration  of a quantum coherent dynamics in a superconducting flux qubit coupled to a superconducting quantum interference device (SQUID) has been reported \cite{chiorescu94}. 

In fact, the coherent dynamics of a quantum system is disturbed by  contact with  the environment \cite{weiss,giulini}. On the one hand, the phase of the quantum coherences is perturbed, which leads to decoherence \cite{jh02}, 
and on the other hand, energy transfer between the system of interest and the environment takes place, which gives
rise to dissipation \cite{weiss,giulini}. 
Besides the fundamental theoretical interest in such quantum dissipative
systems, a deep understanding of the various mechanisms leading to dissipation and decoherence is of
interest in the context of quantum information processing \cite{jh02,thorwart02}. In view of the requirements  expected to be met in the realization of a quantum computer  \cite{divincenzo00},  it is of the utmost importance to control and stabilize the coherent dynamics
of the quantum system of interest. Indeed, various candidates for realizing the building blocks of small scale quantum
information processors with nanoscale solid-state structures have been
proposed \cite{jh03a,jh03b} and also partially realized in ground-breaking
experiments \cite{fazio05}. 

To gain a deeper understanding of the quantum dissipative dynamics, a quantitative
model including the effects of time-dependent driving, decoherence and dissipation is 
required. One generic model to investigate these effects 
is the time-dependent spin-boson model \cite{grifoni98,weiss}, where the tunnel-splitting
of the coherent two-level system (TLS) is usually denoted by $\Delta$. 
The environment is commonly described via the spectral 
density $J(\omega)$ \cite{weiss}. In many cases,  an Ohmic spectral density, 
where $J(\omega)\propto\omega$, occurs, as in the case  of an unstructured 
electromagnetic environment, where all transitions within the system are equally damped \cite{weiss}.

To take into account the fact that the environmental frequencies are in principle limited, a high-frequency cut-off $\omega_c$ has to be introduced. This frequency scale is related to the time scale on which the environmental degrees of freedom evolve \cite{weiss}. In many cases, $\omega_c$ is chosen to be the largest frequency scale in the problem ($\omega_c \gg \Delta$), which corresponds to the fact that the environmental fluctuations evolve on the shortest possible time scale, and hence the bath is `fast'. This describes, e.g., electromagnetic fluctuations in a crystal host \cite{weiss,thorwart05,eckel06} or in a superconductor \cite{weiss}. Under this condition, a Markov assumption can be made yielding an effective time-local dissipative dynamics. This is also the regime in which Bloch-Redfield type master equations are commonly used \cite{may}, typically in connection with the additional assumption of weak system-bath coupling. The opposite case, when $\omega_c\ll \Delta$ describes an effective adiabatic bath which can be treated in a rather simplified manner \cite{weiss}. 
There are, however,  relevant situations when the bath fluctuations occur on a similar time scale on which the system evolves, i.e.,  $\omega_c \simeq \Delta$. This, for instance, typically occurs in the quantum coherent dynamics at the initial stage of 
photosynthesis in complex biomolecular structures  \cite{may,MakriPNAS,Herek02,Brixner05,gilmore06}. It has only recently been experimentally hinted that the efficiency of the energy transfer from the light harvesting
antenna complex to the chemical reaction center is promoted by the appearance of a quantum coherent dynamics \cite{Engel07,Lee07}. This hypothesis is further underpinned in  \cite{Engel07,Lee07}, where it has been evidenced that
the collective long-range electrostatic response of the biomolecular protein environment to
the electronic excitations should be responsible for the measured long-lived quantum coherences. 
In the experiment reported in  \cite{Engel07}, a quantum coherent excitonic dynamics in the energy transfer among bacteriochlorophyll (BChl) 
complexes over a time of around 660 fs has been measured at a temperature of 77 K.
Such a  dynamics 
is highly non-Markovian and more elaborate techniques have to be applied in order to provide  an appropriate theoretical description.           

In what follows, we consider an externally driven TLS subjected to an
Ohmic environment with an exponential cut-off. Here, the focus is on the non-Markovian 
influence of a finite $\omega_c$ on the quantum coherent dynamics. 
The appropriate numerically exact tool to investigate non-Markovian dynamics is 
the quasi-adiabatic propagator path integral (QUAPI)  method, which we briefly describe in section \ref{method}. 
We are able to investigate the entire parameter regime of weak as well as  strong system-bath 
coupling situation beyond the often used scaling limit $\Delta\ll\omega_c$. 
We compare the weak coupling approximation based on 
a Born-Markov approach against  the QUAPI method  and show that the inclusion of non-Markovian
effects is indeed necessary to obtain the correct result in the regime 
$\omega_c \simeq \Delta$ (section \ref{compare}). 
The QUAPI method allows one to go far beyond
the limiting cases within the Born-Markov approach, since
all non-Markovian effects are exactly included. 

 The investigation of the parameter regime $\omega_c \sim \Delta$ is motivated 
by the fact that a slow environment is indeed of physical
relevance to light-harvesting biomolecular 
complexes which are embedded in a polar solvent  \cite{gilmore06}. Here, 
we introduce a model for an effective TLS constructed from two interacting chromophores coupled to a  protein-polar solvent reservoir.  Furthermore, we drive the effective biomolecular TLS with an external laser and show that the Hamiltonian for the full system 
can be described in terms of the driven spin-boson model.  The driven dissipative dynamics  is investigated with the objective of understanding basic quantum interference phenomena  which could be realized as proof-of-principle quantum coherent  control experiments in light harvesting (LH) photosynthetic complexes such as LH II  \cite{may} or in artificially designed TLS nanostructures with specific bath properties.
Only recently, we have demonstrated that in these systems
a non-Markovian environment is most successful in generating entanglement of two non-interacting 
qubits which are coupled to the same Ohmic environment \cite{thorwart08}. 

We  compare our results for the undriven TLS  
with the outcome of real-time Monte
Carlo simulations \cite{muehlbacher03} and show that the QUAPI approach gives reliable results in the 
non-Markovian strong coupling case (section \ref{undriven}).
In addition, we include an external time-dependent drive at frequency $\omega_l$ and show 
in section \ref{driven},  for moderate and strong driving, 
that the amplitude of the forced oscillations in the stationary limit strongly depends 
on $\omega_l$ and, moreover, on $\omega_c$. 
Most interestingly, it turns out that a slow environment together with a slow drive $\omega_l$
optimizes the forced oscillations in the stationary limit. 
Finally, in section \ref{cdt}, the effect of a slow dissipative environment on the coherent destruction of tunneling 
in the TLS is investigated. We find that the bath influence is indeed strongest in the scenario  $\omega_c\simeq\Delta$.

\section{Model and method}

\subsection{Model for the dissipative TLS}
\label{model}
The driven two-level system (TLS) bilinearly coupled to a bosonic heat bath is
described by the generic spin-boson Hamiltonian \cite{weiss}
\beq
H(t)=H_S(t)+H_{SB}+H_{B}\,.
\label{eq:ham}
\eeq
Here, the system Hamiltonian $H_S(t)$ is chosen to be in the basis of
the two states $|0\rangle$ and
$|1\rangle$, each being, for example, the localized charge state
of a charge qubit, or the ground state and the excited state of a two-level
atom. The TLS, with the tunnel splitting $\Delta$, is driven by a
time-dependent external driving field of the form
$\varepsilon(t)=A\cos\left(\omega_lt\right)$ with amplitude $A$ and driving-frequency
$\omega_l$ yielding
\beq
H_S(t)=\frac{\hbar}{2}\left(\Delta\sigma_x+\varepsilon(t)\sigma_z\right)\,,
\label{eq:qubit}
\eeq
with $\sigma_{i=x,z}$ being the Pauli pseudo-spin matrices. The environment
to which the TLS is
bilinearly coupled is modeled as a bath of harmonic oscillators with bosonic creation and
annihilation operators $b^{\dagger},b$ and oscillator frequency
$\omega_{\alpha}$, hence
$H_B=\sum_{\alpha}\hbar\omega_{\alpha}b^{\dagger}_{\alpha}b_{\alpha}$. The
coupling between the TLS and the environment is taken into account by the
interaction Hamiltonian
\beq
H_{SB}=\frac{\hbar}{2}\sigma_z\sum_{\alpha}g_{\alpha}\left(b^{\dagger}_{\alpha}+b_{\alpha}\right)\,,
\label{eq:SB}
\eeq
with $g_{\alpha}$ being the coupling constants.

\subsection{Model for the dissipative photosynthetic light-harvesting effective TLS}
\label{Bmodel}
Complex photosynthetic biomolecular structures have recently been shown to exhibit quantum interference properties \cite{Engel07, Lee07}.  In particular, energy transfer among the excitons within chlorophyll complexes of the sulfur \cite{Engel07} and the purple \cite{Lee07} bacteria have   provided evidence for long-lived (picosecond time scale) quantum coherent excitonic dynamics, a fact that has only recently become associated to the efficiency of the  energy transfer from the LH  antenna complexes to the chemical reaction centres in such large biomolecules \cite{Engel07, Lee07,thorwart08}. 

In this work, we are interested in the dissipative dynamics of the minimal, basic unit of a photosynthetic LH complex which would allow us to model and control  quantum interference mechanisms taking place in such nanostructures. This is done by modeling the specific case of  an interacting pair of  chromophores in a LH II ring  \cite{may}.
The effective single TLS  is built  up from  two chromophores which are coupled by the F\"orster
resonant energy transfer $\Delta$, as sketched in \fref{qubit}(a), where $\hbar\omega_j$ is the transition energy for chromophore $j$.  Since the fluorescence lifetime of the single chromophore is much
larger than the other time scales of the system \cite{gilmore06}, no radiative decay of the excitations is taken into account, and we can write  the Hamiltonian for the two-chromophore  system  in the basis  ${\cal{B}}_2\equiv\{
\left|g_1\ket\otimes\left|g_2\ket,\left|g_1\ket\otimes\left|e_2\ket,\left|e_1\ket\otimes\left|g_2\ket,\left|e_1\ket\otimes\left|e_2\ket\}$,
where $g_j$ ($e_j$) corresponds to the  ground (excited) state of chromophore $j$. 
The  $\Delta$--F\"orster coupling between the two chromophores (figure \ref{qubit}(a)) comprises a dipole-dipole interaction which produces a non-radiative transfer of an excitation between the chromophores.  Such an interaction can be  written as 
$H_{int}=\frac{\hbar\Delta}{2}(\sigma_x^1\sigma_x^2+\sigma_y^1\sigma_y^2)$, and the Hamiltonian of the bare system becomes 
\beq
H_S=H_1\otimes\sigma_0^2+\sigma_0^1\otimes H_2+H_{int}\,,
\label{eq:H_s}
\eeq
where $\sigma_0^i$ is the identity matrix in the space of chromophore  $i$. 

The  correlations due to the bath enter through the coupling to a surrounding protein environment  and to a polar solvent, which, in general, exhibits a frequency dependent dielectric constant \cite{Gilmore08}. For the details of such a mechanism and their possible geometric configurations, we refer to \cite{Gilmore08}. Formally, this process can be modeled by means of a quantized reaction field operator $R^i\equiv \sum_{\alpha}D_{\alpha}^i(b_{i,\alpha}+b_{i,\alpha}^{\dagger})$, where 
 $D_{\alpha}^i$ couples the  chromophores $i$  to the surrounding environment. This,   in turn,  is modeled as a bath of harmonic oscillators which comprise the energy stored in the  polar solvent. Such modes are represented via the
bosonic operators $b_{\alpha}$, and, as in the previous section,  the  bath Hamiltonian reads
$
H_B=\sum_{\alpha}\hbar\omega_{\alpha}b_{\alpha}^{\dagger}b_{\alpha}.
$
If $\delta\mu_i$ denotes the change in the dipole moment of  molecule $i$ during the transition \cite{gilmore06}, i.e., the
difference between the dipole moment of the chromophore in the ground and excited states,
the two chromophores are coupled to their environment via the interaction Hamiltonian
\beqn
H_{SB}=\frac{\hbar}{2}\left[\left(\delta\mu_1\sigma_z^1\sum_{\alpha}D_{\alpha}^1(b_{\alpha}+b_{\alpha}^{\dagger})\right)\otimes\sigma_0^2+\sigma_0^1\otimes \left(\delta\mu_2\sigma_z^2\sum_{\alpha}D_{\alpha}^2(b_{\alpha}+b_{\alpha}^{\dagger})\right)\right]\,.
\nonumber
\eeqn  
The total Hamiltonian for the two-chromophores is then written in the basis   ${\cal{B}}_2$ as
\beqn
\label{eq:H_2C}
H&=&H_S+H_B+H_{SB}=\\
&=&\sum_{\alpha}\hbar\omega_{\alpha}b_{\alpha}^{\dagger}b_{\alpha}+\frac{\hbar}{2}\left(
\begin{array}{cccc}
 -\left(\Omega_++V_+\right) & 0 & 0& 0\\[2mm]\nonumber
0& -\left(\Omega_-+V_-\right) & 2\Delta & 0\\[2mm]\nonumber
0 & 2\Delta & \Omega_-+V_- &0\\[2mm]\nonumber
0 & 0 & 0 & \Omega_++V_+
\end{array}\right)\nonumber\,,
\eeqn 
where $\Omega_{\pm}\equiv\omega_1\pm\omega_2$, and
$V_{\pm}\equiv \delta\mu_1R^1\pm \delta\mu_2R^2$. Given the biophysical nanostructure composition of the LH II rings \cite{may}, we assume that the states of the single chromophores couple to the same surrounding protein bath. Consequently we set $D\equiv D^1=D^2$, and drop any subscripts that may differentiate the bath modes associated to chromophores 1 and 2 in $H_B$\footnote{A coupling of the two chromophores to two uncorrelated baths can easily be 
included within the QUAPI method, but is not within the aim of the present work.}.

If only  singly excited states  are taken into account, from the Hamiltonian \eref{eq:H_2C} we identify the active  environment coupled 2D-subspace $\{|e_1\rangle\otimes|g_2\rangle,|g_1\rangle\otimes|e_2\rangle\}$.  In this central subspace of \eref{eq:H_2C},  the effective interacting biomolecular  TLS 
Hamiltonian reads
\beq
H=\left(\frac{\hbar\Omega}{2}\sigma_z+\hbar\Delta\sigma_x\right)+\frac{\hbar}{2}\sigma_z\sum_{\alpha}g_{\alpha}\left(b_{\alpha}+b^{\dagger}_{\alpha}\right) +\sum_{\alpha}\hbar\omega_{\alpha}b_{\alpha}^{\dagger}b_{\alpha}\ ,
\label{eq:Bqubit}
\eeq
with $g_{\alpha}\equiv D_{\alpha}(\delta\mu_1- \delta\mu_2)$ being the bath coupling constants,  and 
$\Omega\equiv\Omega_-$. Now the defined effective biomolecular TLS has tunneling  splitting $\Delta$; if, as before, such a  TLS  is driven by a
time-dependent external driving field 
$\varepsilon(t)=A\cos\left(\omega_lt\right)$, and we consider that both chromophores have equal transition energies ($\omega_1=\omega_2$), 
the Hamiltonian \eref{eq:Bqubit} becomes equal to the total  Hamiltonian $H(t)$ \eref{eq:ham}, and hence we have effectively mapped the interacting, environment correlated chromophore system Hamiltonian  to that of a  generic  effective spin-boson Hamiltonian.
This is esquematically illustrated in \fref{qubit}(b), where $\Delta$ is the associated ``tunneling energy", between the new basis states $\left|0\ket$ and $\left|1\ket$  for the effective biomolecular TLS (formerly the ${\cal{B}}_2$--basis  states $\left|ge\ket$ and $\left|eg\ket$, respectively).

To gain information on the full dynamics of the system, the initial conditions
at $t=0$ have to be specified. In all our simulations we assume that the full
density matrix $\rho$ at the initial time $t=0$ factorizes in accordance with
\beq
\rho(0)=\rho_S(0)\otimes\rho_B\,.\label{eq:initial}
\eeq
Here, $\rho_S(0)$ is the density matrix of the TLS at initial time $t_0=0$ and
the decoupled bath canonical distribution
$\rho_B=\mbox{Z}_B\exp(-H_B/k_BT)$, where $\mbox{Z}_B=\tr\exp(-H_B/k_BT)$. To keep
the bath in thermal equilibrium, the bath is coupled to a not further
specified super-bath at temperature $T$. 

The environment is fully characterized by the spectral
density $J(\omega)=\sum_{\alpha}g_{\alpha}^2\delta(\omega-\omega_{\alpha})$,
being a quasi-continuous function for typical condensed phase applications. It
determines all bath-correlations that are relevant for the system 
via the bath auto-correlation function \cite{weiss,leggett_rmp}
\beq
L(t)  =   \frac{1}{\pi} \int\limits_0^\infty d\omega J(\omega) \left[ \coth
\frac{\hbar \omega }{2k_BT} \cos \omega t - i \sin \omega t
\right]\,. \label{eq:response}
\eeq
We emphasize at this point that within our numerical method (see below) the full structure of
$L(t)$ is included, which is essential to describe non-Markovian features of
the system dynamics. In particular, we avoid the often used 
Born-Markov approximation, which
consists of replacing the real part of $L(t)$ by a $\delta$-function and neglecting
its imaginary part.

For what is reported in the following, we use an Ohmic spectral density with an
exponential cut-off, i.e., 
\beqn
J(\omega)=2\pi\alpha\omega\exp(-\omega/\omega_c),
\label{eq:ohmic}
\eeqn
where the dimensionless parameter $\alpha$ describes the damping strength and
$\omega_c$ is the cut-off frequency. An Ohmic spectral density is a proper
choice for, e.g. electron transfer dynamics \cite{muehlbacher03,egger94} or biomolecular
complexes \cite{may,gilmore06,thorwart08},  as well as in the case of  Josephson
junction qubits \cite{makhlin01}. In the case of charge qubits
subjected to a phonon bath, a different spectral density, with a super-Ohmic low-frequency behaviour, 
results better suited \cite{thorwart05,eckel06}.

A microscopic derivation for the spectral density  associated to  the bacteriochlorophylls in  the LH II complexes considered in this work  
has been reported in 
\cite{Gilmore08}, where  different forms of a 
Debye dielectric solvent have been  considered. In general, they  lead to the 
Ohmic type of spectral density given by \eref{eq:ohmic}.
The dimensionless  damping  constant $\alpha$ of the protein-solvent 
is directly  related to the parameters of the dielectric model \cite{Gilmore08}, and has been estimated to be  in the range $\alpha \sim 0.01 - 1$ \cite{gilmore06,Gilmore08}.  
The exponential decay of the high-frequency 
cut-off  $\omega_c$ sets the bath characteristic 
time-scale. If
$\Delta \ll \omega_c$, the bath is very fast compared to the
effective TLS and loses its memory quickly. Here,  a Markovian
approximation is appropriate and the standard Bloch-Redfield
description \cite{may} applies. However, for the considered 
biomolecular environment, 
$\hbar \omega_c$ is  typically of the order of  $\sim 2-8$ 
meV, while the F\"orster
coupling strengths $\hbar\Delta \sim  0.2-100$
meV \cite{gilmore06,Gilmore08}. 
Hence, the bath responds slower than the dynamics of the
excitons evolve and non-Markovian effects become 
dominant, a regime  which is accessible only by rather advanced techniques.

Below, we report results in  the scaling limit $\Delta,T\ll\omega_c$ 
and vary $\omega_c$ such that the system reaches the crossover to the 
adiabatic limit, i.e., $\omega_c\sim\Delta$. Both regimes, and the associated crossover, 
have  been studied by real-time Quantum Monte 
Carlo simulations for electron transfer dynamics within the undriven TLS for selected parameter combinations \cite{muehlbacher03}. In this work, we go beyond this by including an external laser driving and, furthermore,  
by covering the entire parameter space. 
Concerning the coupling between the TLS and its environment, we study the whole
parameter window from the weak coupling limit $\alpha\ll 1$  to the strong coupling limit
$\alpha\sim 1$. 

\subsection{The quasi-adiabatic propagator path integral method}
\label{method}
The dynamics of the TLS introduced in the previous section is described in terms of the time
evolution of the reduced density matrix $\rho(t)$ 
which is obtained by tracing over the
bath degrees of freedom. The TLS
dynamics always evolves from the initial state 
$\rho_S(0) =\left|1\ket\bra 1\right|$.

In order to investigate the dynamics of the system, we use the 
QUAPI scheme 
\cite{QUAPI}, which is  a numerically exact iteration scheme which has been successfully 
adopted to many problems of open quantum systems \cite{thorwart05,thorwart08,Tho98}.
For details of the iterative scheme we refer to  previous works \cite{QUAPI,Tho98} and do 
not reiterate them here. 

The algorithm is based on a symmetric Trotter splitting of the short-time
propagator $K(t_{k+1},t_k)$ of the full system into a part depending on $H_S$ and $H_B+H_{SB}$
describing the time evolution on a time slice $\delta t$. This is exact in the limit $\delta t \to 0$
but introduces a finite 
Trotter error to the propagation which has to be eliminated by choosing $\Delta t$ small 
enough that 
convergence is achieved. On the other side, the bath-induced correlations being 
non-local in time 
are included in the numerical scheme over a finite memory time $\tau_{mem}=K\delta t$ 
which roughly corresponds to the time range over which 
the bath auto-correlation function $L(t)$ given in \eref{eq:response}  
is significantly different from zero. As the environmental fluctuations live on a time scale 
$\sim 1/\omega_c$, it is particularly important to include the full memory when 
$\omega_c \simeq \Delta$. Note that for any finite temperature, $L(t)$ 
decays exponentially at long times \cite{weiss}, thus justifying this approach. Moreover, 
$K$ has to be increased, until convergence with respect to the memory time has been found. 
Typical values, for which convergence can be achieved for  
our  spin-boson system, are $K\leq 12$ and $\delta t \sim (0.1 - 0.2) / \Delta$. 
The two strategies to achieve convergence are countercurrent. To solve this, 
the principal of least dependence 
has been invoked \cite{Tho98} to find an optimal time increment in between the two 
limits $K\to \infty$ and $\delta t \to 0$.

\section{Dynamics of the driven TLS}

With the time evolution of the reduced density matrix $\rho(t)$ at hand
we can now study the dynamics of the driven TLS in terms of  the population difference 
$P(t)=\langle \rho(t) \sigma_z\rangle$ of
the two states, with the initial condition $P(0)=1$.

\subsection{Markovian vs non-Markovian dynamics}
\label{compare}
Before addressing the effect of driving,  we convince ourselves that the dynamics is 
indeed non-Markovian when $\omega_c \simeq \Delta$. For this, we 
compare the numerical exact QUAPI with a Bloch-Redfield approach. 
To be specific, we compare the QUAPI data with the outcome
of the weak coupling approximation, see (21.171) and (21.172)
in \cite{weiss}, which results from a first order approximation in $\alpha$.  
In \cite{hartmann00}, it has been shown that the outcome of the
weak coupling approach is equivalent to a Bloch-Redfield treatment.

In \fref{wca1} (a) the result for $\alpha=0.001$ and $\omega_c=100\Delta$ is shown. 
As expected, the agreement between both results is perfect since the system is 
deep in the Markovian (weak coupling) regime. For \fref{wca1} (b) the coupling
is increased to $\alpha=0.01$ and since the bath is still in the scaling limit
there is acceptable agreement. In contrast, for a cut-off frequency $\omega_c=\Delta$,
see \fref{wca1} (c), strong deviations between the Bloch-Redfield approach and
the QUAPI approach arise, indicating that the environment is non-Markovian
in nature here. 
Although the smooth cut-off function with the reduced spectral weight at frequencies $\omega\sim\omega_c>\Delta$ 
is fully included within the Bloch-Redfield approach, we see strong deviations. 
This further underpins that, as expected, the Bloch-Redfield approach fails 
when the bath fluctuates on a time scale comparable to that of the system's time evolution. 

When the damping is furthermore increased to $\alpha=0.1$ the
disagreement between the numerical exact QUAPI and the weak coupling approximation
is enhanced, see figs. \ref{wca2} (a) and (b). This indicates that the choice of
$\omega_c$ away from the scaling limit induces a non-Markovian behaviour of the 
dissipative TLS dynamics which is further underpinned by \fref{wca2} (c), where the
relaxation rate $\gamma_r$ is shown. We obtain this from a fit of a decaying cosine function with a 
single exponential to $P(t)$. In the scaling limit $\Delta\ll\omega_c$ both
approaches yield the same $\gamma_r$, whereas strong deviations occur for smaller $\omega_c$. 
Note that $\omega_c$ enters (21.171) in \cite{weiss} via the renormalized tunneling
amplitude $\Delta_{\rm eff}$ which is included here in the weak-coupling approximation. 

\subsection{Undriven case: comparison with Quantum Monte Carlo results}
\label{undriven}
Next, in order to validate our results in the highly non-Markovian
crossover regime $\omega_c\sim\Delta$, 
we compare our results with the outcome of real-time Quantum Monte
Carlo (QMC) simulations \cite{egger94}. 
In \fref{fig1} (main), the  results for the high temperature regime
are shown, where the parameters are $T=4\omega_c$ and
$\alpha=2$ (strong coupling). In perfect (also quantitative) 
agreement with \cite{egger94}, $P(t)$ decreases faster for
$\Delta/\omega_c=2.4$ than for $\Delta/\omega_c=1.6$. 
The cut-off frequency is also related to the 
reorganization energy of the environment \cite{weiss}, which has the form   
$E_r=\int_0^{\infty}d\omega\frac{J(\omega)}{\pi\omega}=2\alpha \hbar \omega_c$
for an Ohmic environment. Hence, our results are consistent with the
physical expectation, since the dynamics of the environment is slower for
the smaller ratio $\Delta/\omega_c$ for the same $\alpha$. 
For the low temperature regime $T=0.4\omega_c$, shown in the inset of \fref{fig1}, we also observe
agreement of the QUAPI results with the outcome of \cite{egger94}.

Furthermore, we address the case of strong coupling between the TLS and its
environment. For the scaling limit
$\Delta\ll\omega_c$ it is known that there is a transition at $\alpha=0.5$
from a coherent decay of $P(t)$ for $\alpha< 0.5$ to an incoherent decay
for $\alpha> 0.5$ \cite{weiss,egger97}. To be specific, we choose $\alpha=0.6$ at a
low temperature $T=0.1\Delta$, as shown in \fref{fig2}. Again, this behaviour can be
understood in terms of the reorganization energy $E_r$, since decreasing
$\omega_c$ here has a similar effect as lowering the damping parameter
$\alpha$. By means of our numerical exact method, \fref{fig2} quantifies how
the transition between coherent and incoherent behaviour depends on $\omega_c$
for a given $\alpha$.  

We next  put our results in the context of the modeled effective biomolecular  TLS.  
For excitations in the LH-II ring of the bacteria chlorophyl molecule (BChls in LH-II complexes), the F\"orster
coupling strength $\hbar\Delta \sim  46-100$ meV, and $\hbar \omega_c\sim 2-8$ 
meV   \cite{gilmore06,Gilmore08}, and hence the ratio $\frac{\omega_c}{\Delta}\sim 0.1$.  For these complexes, $\alpha$ is of the order of $0.1-1$ \cite{Gilmore08}, evidencing a strong coupling  between the chromophores and the solvent dielectric. If we make  $\alpha=0.1$, $\hbar\Delta =100$ meV, and plot a graph such as the one of  \fref{fig2} for $T=0.1\Delta$ ($\sim 116$ K), long-lived coherent oscillations are sustained for a time of around  530 fs ($t\Delta\sim 80$). Such oscillations can be visualized at fixed    $\omega_c=0.1\Delta$ (not shown). This rough estimation is in agreement with the time scale of the coherent oscillations recently measured in \cite{Engel07} for the antenna complex from a green sulphur bacteria that has seven BChls per protein subunit. If, on the other hand, we make $\alpha=0.6$,   $\omega_c=0.1\Delta$ (as shown in 
\fref{fig2}), coherent oscillations are also found but, due to the stronger coupling to the bath, they decay  quicker than for the case $\alpha=0.1$. We thus observe that  for the rather simple effective biomolecular TLS model introduced here, which is aimed as a guide to  the  possible  realization of further proof-of-principle experiments,
we are able to  demonstrate that the non-Markovian features of the protein-solvent environment  help to sustain the quantum coherence mechanisms exhibited by the coupled chromophores in a LH-II ring. Since our results are of a general character, in principle derived from a generic TLS, we also expect them to be valid in artificially designed nanostructures with the specific bath properties described here.  

\subsection{Driven case $A\neq 0$}
\label{driven}
We now address the influence of a finite periodic external driving field. 
The results for $A=\Delta$ and $\omega_l=0.05 \Delta$ are shown in  
\fref{fig3}. Similarly to the undriven case, the overall decay of $P(t)$
is faster for larger $\omega_c$:  compare, for example, \fref{fig3} (a) 
for $\omega_c=1.5\Delta$ and \fref{fig3} (b) for $\omega_c=30\Delta$, both in the
weak-coupling situation $\alpha=0.01$. In turn, for the case
$\omega_c\sim\Delta$, the superimposed oscillations due to coherent tunneling 
survive longer than in the scaling-limit $\Delta\ll\omega_c$ (\fref{fig3}(b)), before
the system reaches its stationary state. There, only the stationary oscillations due to 
the external driving field survive.
For a stronger system-bath coupling $\alpha$ the decay of $P(t)$ is faster, as expected,  
and the stationary state is reached faster (not shown). As in the undriven case, the described behaviour
is qualitatively understandable in terms of the reorganization energy $E_r$. 

For increasing driving strength to $A=10\Delta$, the dependence of the TLS dynamics
on $\omega_c$ and $\alpha$ is similar. However, the form of the stationary oscillations 
turns out to be qualitatively different from the case of weak 
driving $A=\Delta$. Here, stable stationary plateaus emerge, as shown in \fref{fig4}. 
In \cite{noel98}, this has been observed experimentally for frequency-modulated
excitations of a two-level atom, using a microwave field to drive transitions between two Rydberg-Stark states of
potassium. In the presence of a slow frequency modulation, square wave oscillations of
the population difference have been detected. 
They can be understood to mean that the large driving amplitude
leads to an extreme biasing of the TLS dynamics. 
The centres of the observed plateaus correspond to the extrema of
the applied cosine driving field. At the position of these maxima, the TLS is maximally biased 
and since the time-scale
of the driving field is much smaller than the time-scale of the (unbiased) TLS dynamics 
due to $\Delta$, the situation of an extreme quasistatic
bias results, forming an intermediate self-trapping around the maxima 
of the cosine-like driving field. Indeed, this intermediate self-trapping becomes
shorter lived, and increasingly washed out, when the $\omega_l\sim\Delta$ (not shown here). 
For smaller $\alpha$ (\fref{fig4}a), the square wave form is superimposed by fast ''intrawell'' damped 
oscillations which die out on a time scale $\sim 1/\alpha$. For increasing $\alpha$  
(\fref{fig4}b), the fast ``intrawell'' oscillations are not present anymore and stable plateaus form immediately 
after the ``interwell'' transition. Due to the strong damping, the level is also strongly broadened and the 
quasistationary picture applies in which the level position is given by the mean energy of the broadened level. 
This is not so much sensitive anymore to further variation within a half period of the cosine field.

In a next step, it is interesting to consider the amplitude $A_{\infty}$ 
of the forced oscillations in the stationary limit. In \fref{fig5}, we show the 
results for the nonlinear response for the case $\alpha=0.1$ and a 
driving field amplitude $A=\Delta$. We observe a rather weak dependence on $\omega_l$ when 
$\omega_l \lesssim \Delta$. Moreover, the response also becomes rather weak in 
the regime of strong detuning. In between these two regimes, 
we find an optimal driving frequency $\omega_l\sim \Delta/2$, where the amplitude of the
forced oscillations has a maximum. Note that this rather weak resonance is due to the sizeable 
damping $\alpha=0.1$. The maximum can be related to the picture of quantum stochastic resonance in an unbiased but strongly damped 
bistable system \cite{Thorwart97}. The response of the system becomes optimal when the dissipative 
tunneling transitions are in synchrony with the external drive. In addition to this well known mechanism, we study here 
the dependence on the bath cut-off frequency. The behaviour depicted in \fref{fig5} is, essentially, 
not influenced by $\omega_c$ for this value of the driving strength $A$ and 
is thus independent of the time scale of the environment (up to small 
quantitative modifications). 

A natural question is whether to expect an enhancement of the response when all frequencies are comparable, i.e., 
$\omega_l \sim \omega_c \sim \Delta$. This situation is addressed in \fref{fig6}, where 
we have chosen $\omega_l=0.5 \Delta$. We find that for weak to 
intermediate driving, no pronounced resonance appears, as shown in  \fref{fig6} (a) for the case
$A=\Delta$. In contrast, strong driving can induce a resonant nonlinear response which, however, requires 
non-Markovian dynamics, i.e., $\omega_c \sim \Delta$. This is illustrated in \fref{fig6} (b) for $A=10\Delta$. 
A slow non-Markovian bath with $\omega_c\lesssim\Delta$ is
thus much more efficient in maximizing forced oscillations in the stationary limit. 
This feature occurs for the weak coupling ($\alpha=0.01$) as well as for the strong
coupling case $\alpha=0.6$. This constitutes another example where a non-Markovian bath helps in increasing coherence 
in the quantum system \cite{thorwart08}. 

In contrast, a Markovian environment in the scaling limit, 
$\Delta\ll\omega_c$, largely suppresses forced oscillations via its destructive influence on 
coherence. This finding 
 is most pronounced in the incoherent strong coupling case, $\alpha=0.6$. 
The dependence of $A_{\infty}$ on $\omega_c$ is again qualitatively understandable via the 
reorganization energy $E_r$ in the case of weak driving. It explains the reduction of the response  
for weaker damping ($\alpha=0.1$ and $\alpha=0.01$), compared 
to the strong damping case $\alpha=0.6$ (\fref{fig6} (a)). However, for stronger driving, 
$A=10\Delta$, the resonance  effect is more pronounced 
for a strong coupling situation, $\alpha=0.6$ which cannot be explained in terms of 
a growing reorganization energy $E_r$. 

\subsection{Coherent destruction of tunneling} 
\label{cdt}
When an isolated symmetric quantum TLS is driven with large frequencies 
$\omega_l\gg\Delta$, the bare tunneling matrix element 
effectively becomes renormalized by the zero-th Bessel function $J_0(x)$ as 
 $\Delta\rightarrow J_0(A/\omega_l)\Delta\equiv\Delta_{\rm eff}$ 
\cite{grifoni98,shirley65}. The population 
of the state $\left|1\ket$ 
with an initial preparation $\left|1\ket=1$ follows as 
\beq
P_1(t)=\cos^2\left(J_0(a/\omega_l)\Delta t/2\right)\,.
\eeq     
This implies that for particular choices of the driving parameters, 
the Bessel function term can be fixed to zero.  
The first zero then corresponds to $A/\omega_l=2.40482...$, and then,  
$P_L(t)$ equals unity, since the effective tunnel splitting vanishes. This phenomenon is
known as coherent destruction of tunneling (CDT); see \cite{grifoni98}
and references therein for further details.

Naturally, the phenomenon of CDT  is influenced  
when the TLS is coupled to an Ohmic environment. A complete standstill of the dynamics will 
no longer occur, due to the relaxation processes induced by the bath. 
For an Ohmic environment in the scaling limit under the assumption of 
weak damping, this question has been addressed 
in \cite{thorwart00}. Here, the QUAPI technique allows us to extend these studies to the case of 
finite $\omega_c$. To be specific, we choose a driving frequency $\omega_l=20\Delta$. 

For weak coupling, $\alpha=0.01$, 
the CDT is only weakly influenced by the environment, as expected, and it turns out
that the dependence on the cut-off frequency of the bath $\omega_c$ is also weak, 
as shown in  \fref{fig7} (left). Nevertheless, we find that the unavoidable decay of $P(t)$ 
occurs the slowest when $\omega_c \sim \Delta$. 
For stronger coupling $\alpha=0.1$, as shown in  \fref{fig7} 
(right), CDT is more strongly influenced by the magnitude of $\omega_c$. 
We observe, again, that a slow bath helps 
to preserve coherence and the decay of $P(t)$ is slow. 

In the regime of strong damping, the situation is different. 
In fact, we find an opposite qualitative behaviour 
 which goes beyond the above given explanation in terms of $E_r$. 
In \fref{fig8}, the corresponding results for $\alpha=0.6$ are shown. 
As one can see, the decay of $P(t)$ is the fastest when 
$\omega_c \simeq \Delta$. This stems from the fact that
for $\omega_c\sim\Delta$, non-Markovian effects are not negligible, since 
the time-scales of system and environment are of the same order, which is reflected in the fact that 
the spectral density $J(\omega)$ has a maximum around $\Delta$. In this resonant situation,
the bath modes
around the characteristic time scale $1/\Delta$ of the TLS are most important and 
disturb the CDT in a maximal (resonant) way. Consequently, the
decay under the CDT conditions is most pronounced.  
In turn, in the scaling limit, the decay under CDT 
conditions is slow. Interestingly, in both cases of strong coupling ($\alpha=0.1;0.6$)  the population difference is almost  maintained on the considered time windows 
without decay by the  slow bath $\omega_c\sim 0.1\Delta$, and the decay of $P(t)$ is the slowest. 
          
\section{Conclusion}
\label{concl}
We have investigated the dynamics of the driven
spin-boson system in the presence of an Ohmic bath. The focus is put on the role of the cut-off frequency 
$\omega_c$. Based on the numerical exact QUAPI approach, non-Markovian effects are shown to be relevant 
 when $\omega_c\sim\Delta$. This effect is more pronounced for  strong damping, as expected, and as can be seen
from the relaxation rate $\gamma_r$ shown in \fref{wca2}. 
The validity of the
QUAPI method in the regime $\omega_c\sim\Delta$ is confirmed by  the 
perfect agreement with published results of real time Quantum Monte 
Carlo simulations \cite{muehlbacher03}.

For the unbiased case and for  strong coupling, we show that damped coherent oscillations exist  in the population difference $P(t)$ if the
bath has a cut-off frequency away from the scaling limit, where the decay is
known to be incoherent in nature, as shown in \fref{fig2}. 
By comparison with relevant experimental data, we were able to
show that our results are directly applicable to biomolecular systems, namely to the light harvesting complex LH-II of the bacteria chlorophyl molecule. We have shown that a sub-unit of such a biomolecular system can effectively be described by means of the spin-boson model and have demonstrated that the non-Markovian features of the protein-solvent environment  help to sustain the quantum coherence mechanisms exhibited within the LH-II complex.
Moreover, since our results are based on a model of general character, we expect them to apply also for a variety
of artificially designed nanostructures with the specific bath properties reported here.

Regarding the coherent control of the effective TLS, we found that 
a strong external driving amplitude in combination
with a slow driving laser frequency produce a population difference with square wave oscillations
in the stationary limit, in agreement with the experimental results  reported in \cite{noel98}.
These square-wave like oscillations stem from the fact that 
the TLS experiences a large quasistatic bias.  
Moreover, it was shown that the amplitude $A_{\infty}$ 
of the forced oscillation in the stationary limit depends strongly on the
frequency of the driving field. A slow driving $\omega_l\lesssim\Delta$ 
field protects the forced oscillations against the influence of the dissipative
environment, whereas in the case of a faster driving, these oscillations are 
strongly suppressed. 
This is valid for the weak-coupling as well as
for the strong coupling case. For larger driving frequencies, the forced
stationary oscillations are considerably reduced. 
The stationary amplitude of the forced oscillations also
strongly depends on the time-scale of the environment determined by $\omega_c$. For strong driving, 
we find that the stationary amplitude shows resonant features, illustrating that the 
non-Markovian environment plays a constructive role. 

Finally, we investigated  the phenomenon of coherent 
destruction of tunneling under the bath  influence.  Away from the scaling limit, 
the influence of the environment is weaker and the CDT survives on a significantly longer
time scale. For very strong coupling, i.e., deep in the incoherent regime, the situation is completely different.
Here, the CDT is most quickly destroyed when the time scales of the TLS and the environment are in resonance.  In such a strong coupling scenario, the preservation of coherence survives longer in the scaling limit; moreover, the CDT is actually helped by the effect of a slow bath (e.g., $\omega_c\sim 0.1\Delta$), where the decay of $P(t)$ is the slowest.

\ack
J.E.\ wishes to thank the Departamento de F\'isica of the Universidad del Valle in Cali (Colombia)  for the kind 
hospitality during his stay, where parts of this work were completed. 
This work was supported by the DFG Priority Program 1243, by the Excellence Initiative 
of the German Federal and State Governments, the DAAD-PROCOL
Program, and by Colciencias grant 1106-452-21296. Computational time of the ZIM at the Heinrich-Heine-Universit\"at
is also acknowledged.

\bfig[h]
\bc
\includegraphics[width=12cm]{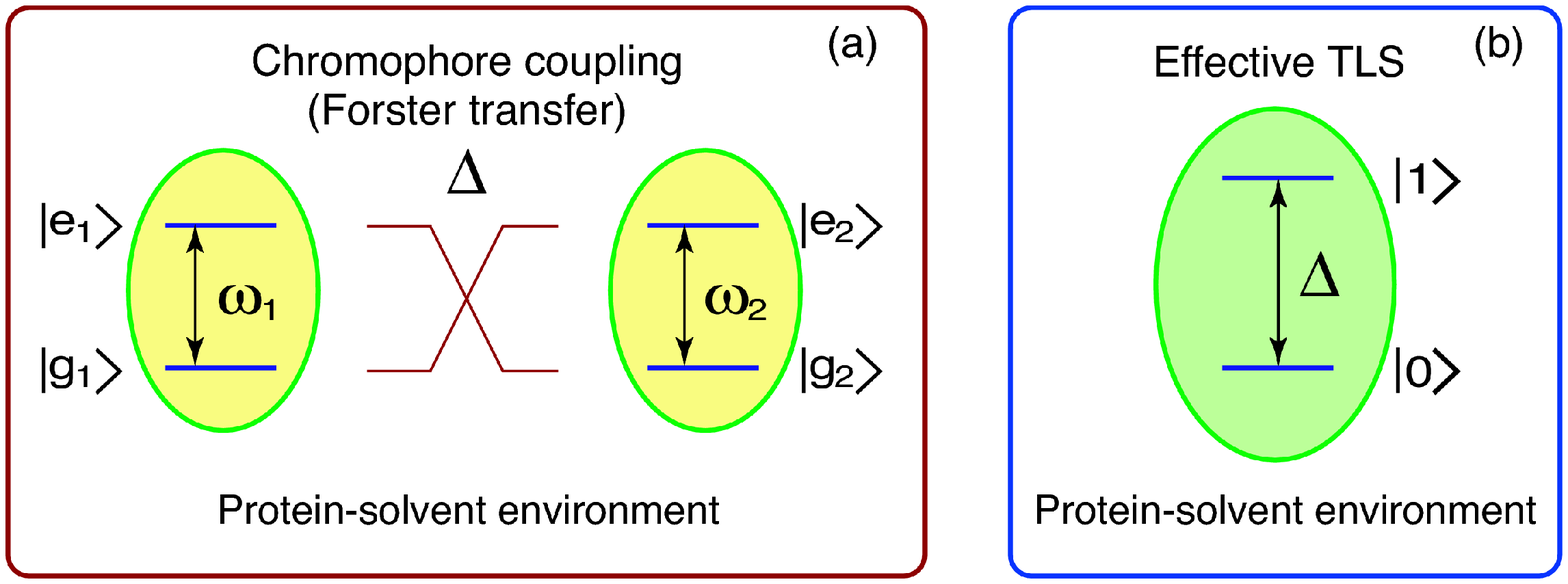}
\caption{Sketch of the effective light harvesting biomolecular TLS formed from a pair of $\Delta-$F\"orster interacting chromophores. \label{qubit}}
\ec
\efig

\bfig[h]
\bc
\bmp{8.0cm}
\includegraphics[width=8cm]{fig2a.eps}
\emp
\bmp{8.0cm}
\includegraphics[width=8cm]{fig2b_2c.eps}
\emp
\caption{Population difference $P(t)$ for the undriven spin-boson model $A=0$ (solid line: QUAPI, symbols: weak coupling approach). (a)
$\alpha=0.001$ and $\omega_c=100\Delta$ (Markovian regime), (b) $\alpha=0.01$ and
$\omega_c=100\Delta$, and (c) $\alpha=0.01$ and $\omega_c=\Delta$. The temperature
is always $T=10\Delta$.\label{wca1}}
\ec
\efig

\bfig[h]
\bc
\bmp{8.0cm}
\includegraphics[width=8cm]{fig3a_3b.eps}
\emp
\bmp{8.0cm}
\includegraphics[width=8cm]{fig3c.eps}
\emp
\caption{Same as \fref{wca1}, but for a strong coupling $\alpha=0.1$. (c) shows the 
relaxation rate $\gamma_r$ as a function of $\omega_c$.  \label{wca2}}
\ec
\efig

\bfig[h]
\bc
\includegraphics[width=9cm]{fig4.eps}
\caption{Population difference $P(t)$ for the TLS for different cut-off frequencies
$\omega_c$ and a damping parameter $\alpha=2$. 
The temperature $T=4\omega_c$ (main), and $T=0.4\omega_c$ (inset).\label{fig1}}
\ec
\efig

\bfig[h]
\bc
\includegraphics[width=9cm]{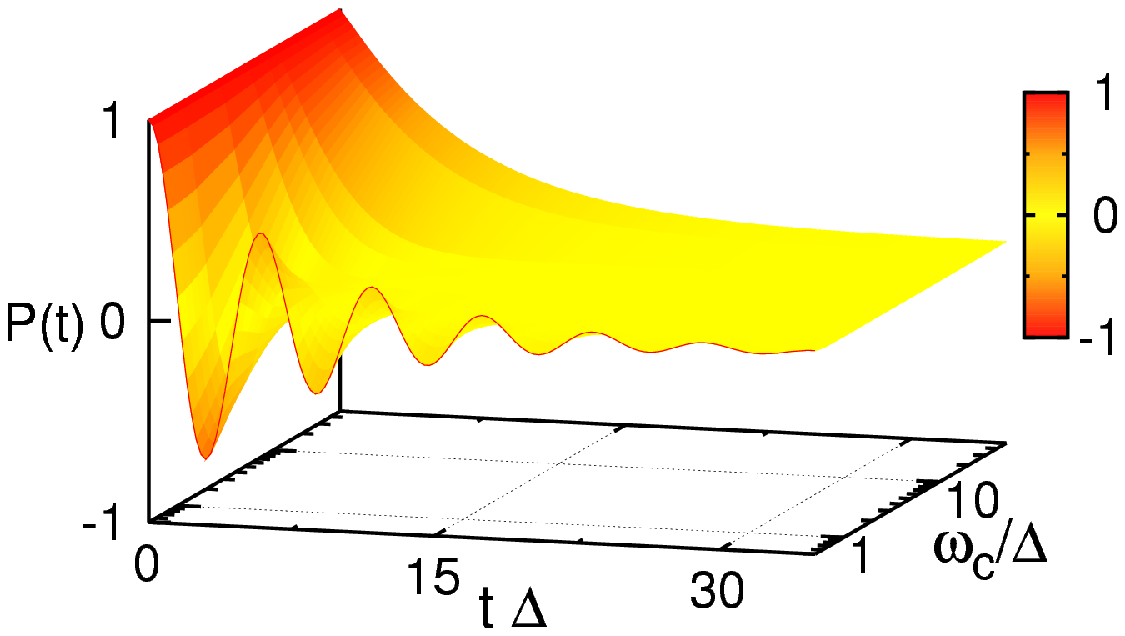}
\caption{Population difference $P(t)$ for the undriven TLS system in
  dependence on the cut-off frequency $\omega_c$. The
  temperature  $T=0.1\Delta$, and the damping parameter is $\alpha=0.6$. \label{fig2}}
\ec
\efig

\bfig[h]
\bc
\includegraphics[width=9cm]{fig6a_6b.eps}
\caption{Population difference $P(t)$ for the driven TLS. The
  amplitude of the driving-field is $A=\Delta$ and the driving-frequency is $\omega_l=0.05\Delta$.
  The temperature $T=0.1\Delta$, and the damping parameter $\alpha=0.01$. The cut-off frequency is
  (a) $\omega_c=1.5\Delta$ and (b) $\omega_c=30\Delta$\label{fig3}}
\ec
\efig
\bfig[h]
\bc
\includegraphics[width=9cm]{fig7a_7b.eps}
\caption{Population difference $P(t)$ for the driven single qubit system. The
  amplitude of the driving-field $A=10\Delta$, the driving-frequency $\omega_l=0.05\Delta$, the cut-off frequency
$\omega_c=0.5\Delta$, 
  and the temperature $T=0.1\Delta$. The damping parameter is (a) $\alpha=0.1$, and (b) $\alpha=0.6$. \label{fig4}}
\ec
\efig
\bfig[h]
\bc
\includegraphics[width=9cm]{fig8.eps}
\caption{Amplitude of the forced oscillations in the stationary limit $A_{\infty}$ as a function of the driving frequency $\omega_l$, for $A=\Delta$ with $\alpha=0.1$ 
and different cut-off frequencies $\omega_c$.\label{fig5}}
\ec
\efig
\bfig[h]
\bc
\includegraphics[width=9cm]{fig9a_9b.eps}
\caption{Amplitude  $A_{\infty}$ of the forced  stationary oscillations 
 as a function of the cut-off frequency $\omega_c$ for (a)  weak to intermediate driving 
$A=\Delta$, and (b) strong driving $A=10 \Delta$, for three different values of $\alpha$. The remaining parameters are 
$\omega_l=0.5 \Delta$, and $T=0.1\Delta$. 
\label{fig6}}
\ec
\efig

\bfig[h]
\bc
\bmp{8.0cm}(a) 
\includegraphics[width=8cm]{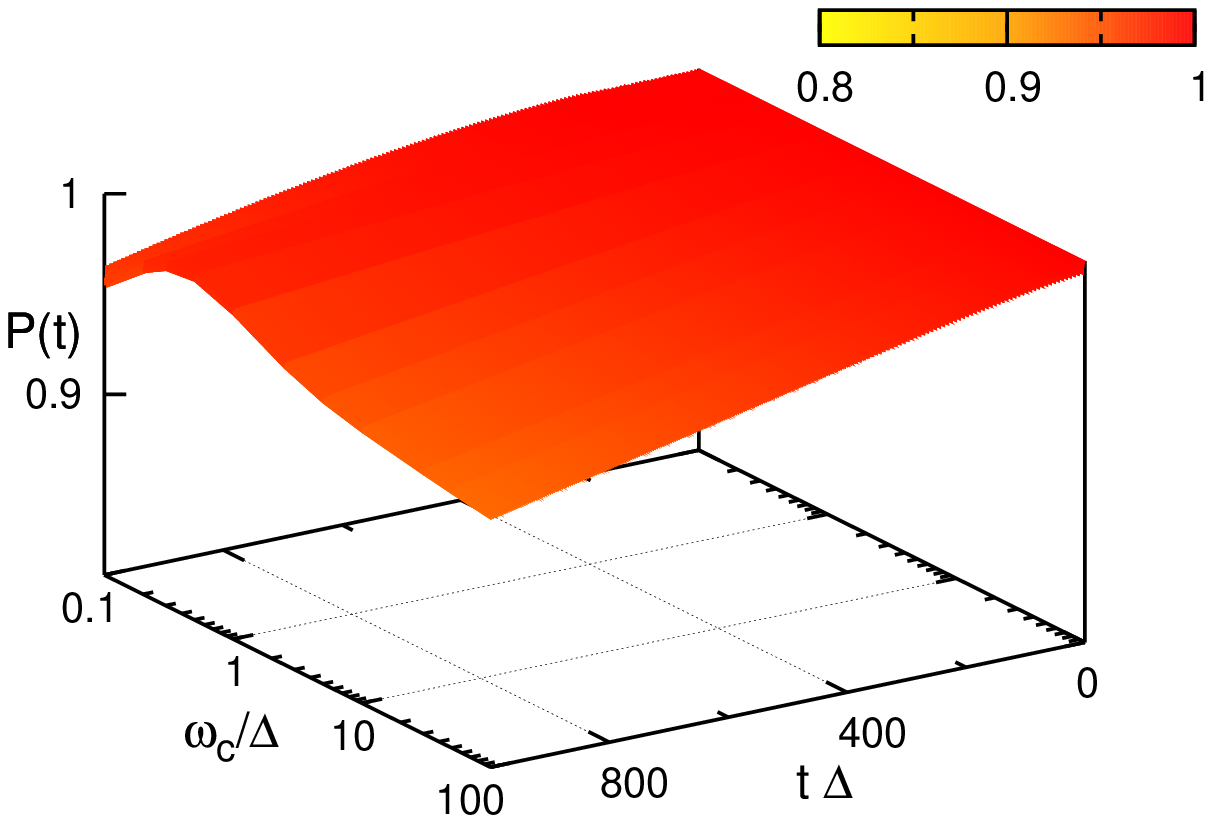}
\emp
\bmp{8.0cm}(b) 
\includegraphics[width=8cm]{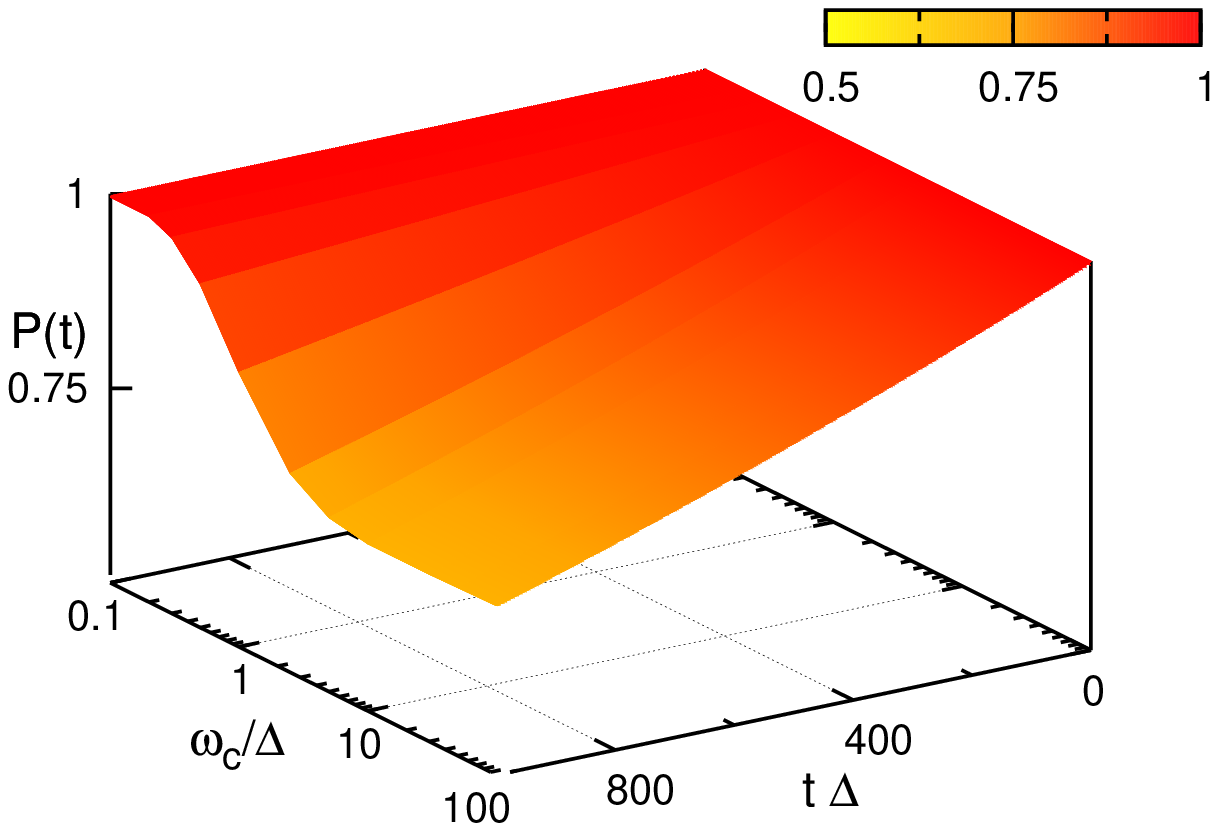}
\emp
\caption{Population difference $P(t)$ for the driven TLS 
under the condition for coherent destruction of tunneling, $\omega_l=20\Delta, A= 
2.40 \Delta$. The temperature $T=0.1\Delta$,  
and the damping is (a) $\alpha=0.01$, and (b) $\alpha=0.1$. \label{fig7}}
\ec
\efig

\bfig[h]
\bc
\includegraphics[width=9cm]{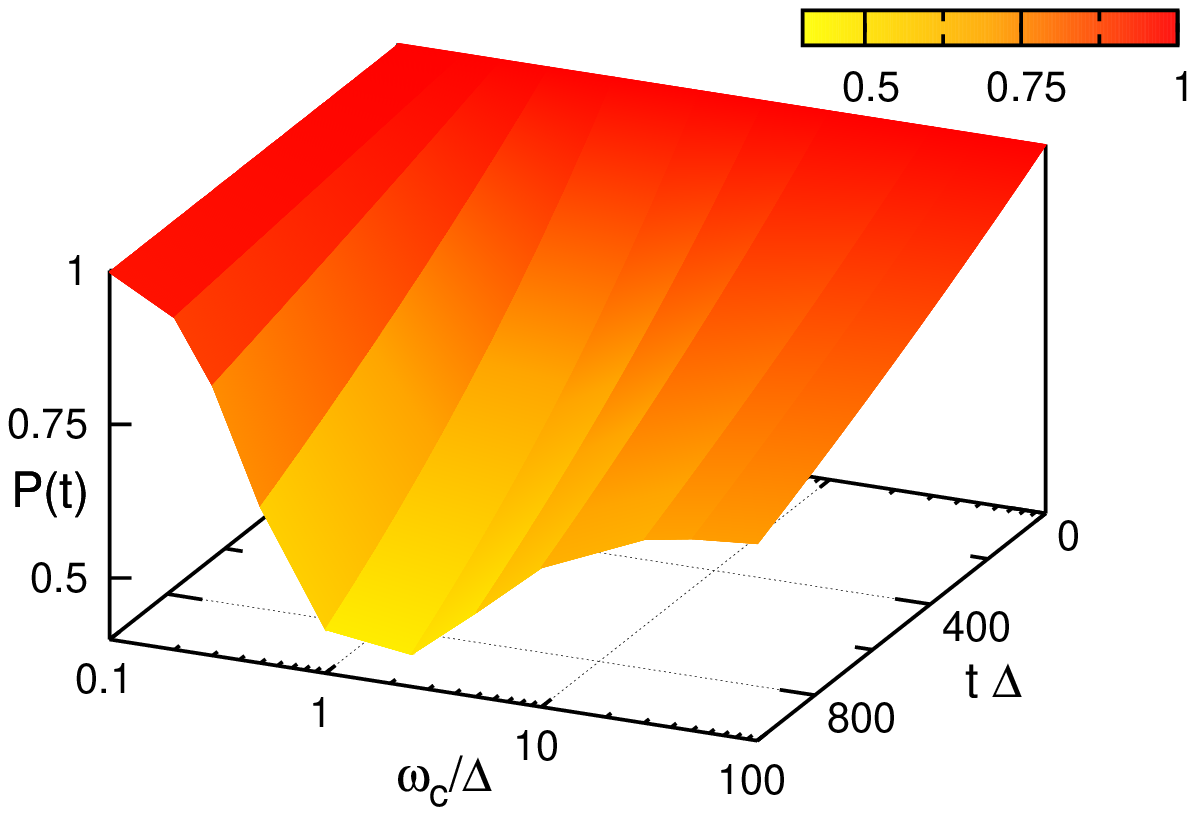}
\caption{Same as \fref{fig7}, but for a strong coupling situation, $\alpha=0.6$. \label{fig8}}
\ec
\efig

\end{document}